# Performance Measurement of the Heterogeneous Network

**Soumen kanrar  and Mohammad Siraj**

Department of Computer Engineering,
College of Computer and Information Sciences
King Saud University - Riyadh

**Abstract**

Till today we dreamt of imperceptible delay in a network. The computer science research grows today faster than ever offering more and more services (computational representational, graphical, intelligent implication etc) to its user. But the problem lies in *"greater the volume of services greater the problem of delay"*. So tracing delay, or performance analysis focusing on time required for computation, in a existing or newly configured network is necessary to conclude the improvement. In this paper, we have done the job of delay analysis in a multi-server system,. For this proposed  work we  have used continuous –parameter Markov chains (Non –Birth –Death Process)**,**for developing the required models, and for developing the simulator we have used queuing networking, different scheduling algorithms at the servers queue and process scheduling .
The work can be further extended to test the performance of wireless domain.

## Problem definition

Define the problem in the following ways.
let G be a closed and connected graph such that
G=(V,E) where V is the set of vertices or nodes,
V={ $v_k$ | k=1,2…,n}

and E is the set of edges, E={ ( $v_k$ , $v_j$ )| k $\neq$ j

and k =1,2,..,n
and j=1,2….,n}
and n is the number of nodes in the graph.
Now we are considering a closed , directed and connected graph G such that –

( $v_k$ , $v_j$ ) $\notin$ E  $\forall$  k= 1,2,..,(n-3)  $\wedge$  j =1,2,…,(n-3)

( $v_k$ , $v_j$ ) $\notin$ E  $\forall$  k= 1,2,..,(n-3)  $\wedge$  j =(n-1),n

( $v_k$ , $v_j$ ) $\notin$ E   for k= (n-1)  $\wedge$  j =n

( $v_k$ , $v_{n-2}$ ) $\in$ E  $\forall$  k= 1,2,..,(n-3)

( $v_{n-2}$ , $v_j$ ) $\in$ E  $\forall$  j= (n-1),n

To measure the delay at the node $v_{n-2}$  when nodes $v_k$ , k=1,2,..(n-3) send the jobs and the jobs are executed at the nodes  $v_{n-1}$ and $v_n$ .

## The actual problem

we can view the problem of delay analysis in a Network as the problem stated above. We can imagine that $v_1, v_2,...., v_{n-3}$ are the clients those are sending jobs at unpredictable interval of unpredictable volume. The jobs form a queue for waiting at the node   $v_{n-2}$ ( an intermediate machine) and then are sent to the server $v_{n-1}$ and $v_n$ for service. We offer the responsibility of maintaining the queue and dispatching them to the server depending on their capability. Now we are required to find the queuing delay at the queue, transmission delay, processing delay etc. at the intermediate machine (node $v_{n-2}$ ) the problem becomes more realistic and logical.

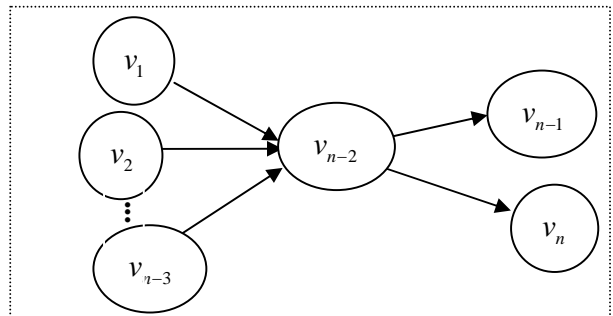

**Graphical Representation of the Problem**

## So we can state the problem as

The clients C1, C2, ….,Cn are sending their jobs with different probabilities $\lambda_1, \lambda_2,..., \lambda_{n-1}$ and they forming a queue under the supervision of intermediate
node ( I ) and then dispatched to the servers S1 and S2 with probability  $\mu_1$ and

$\mu_2$ depending on the capability of the servers to service. In the intermediate machine (I)
we did our calculations for performance measurement .

---

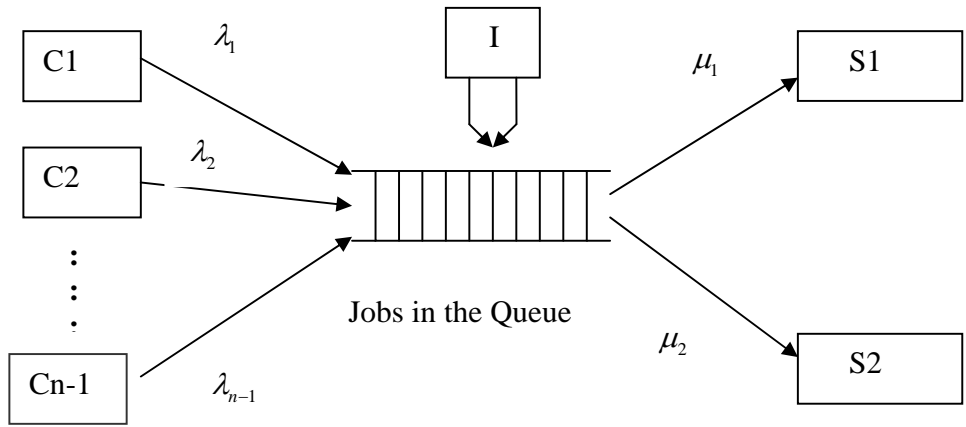

**Queue with heterogeneous servers**

In an M/M/2 queuing system with heterogeneous servers there are two server , which are of different processing capability i.e service rates of the two servers are not identical . The discipline used to schedule the jobs is FCFS. This is equivalent to saying that jobs are served in the order of their arrival.

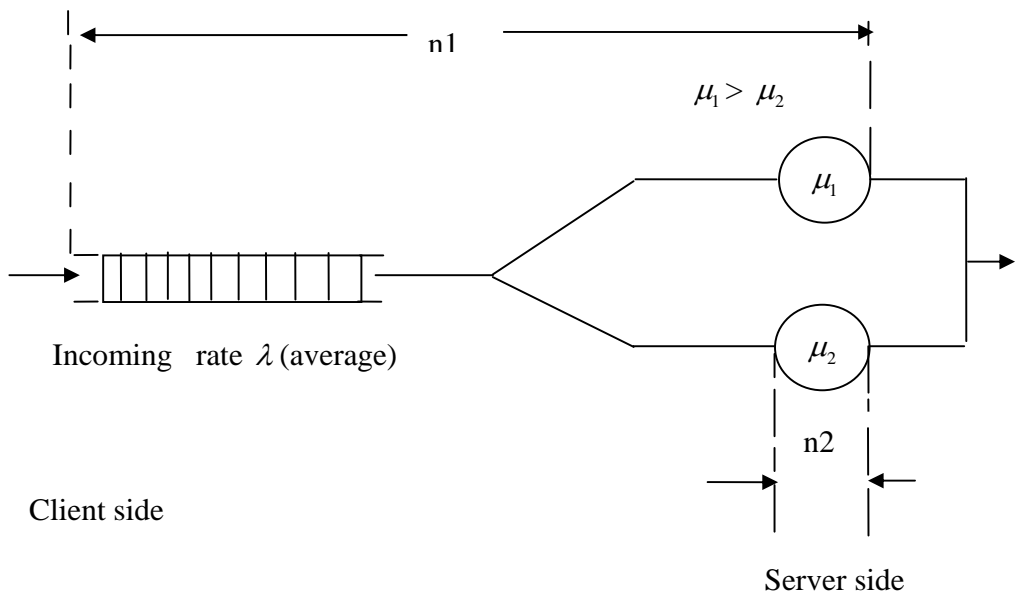

**Heterogeneous System**

The state of the system is defined to be the tuple (n1,n2) where n1 denotes the number of Jobs in the queue including any at the faster server ,and n2 denotes the number of jobs at the slower server .Jobs wait in the order of their arrival .When both servers are idle, the faster server is schedule for service before the slower one.

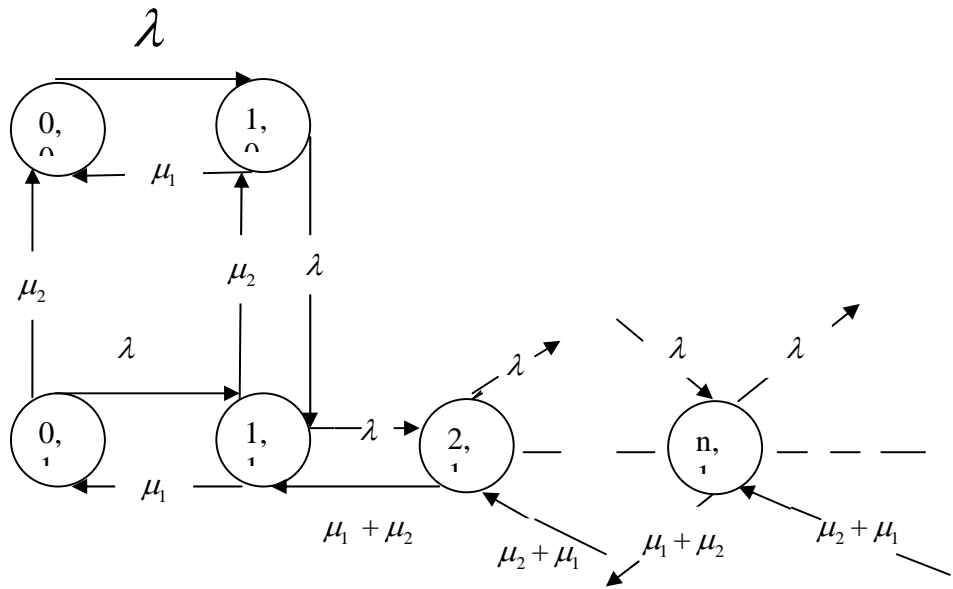

**State Transition Diagram of Heterogeneous system**

In the steady state we have

$\lambda\, p(0,0) = \mu_1 p(1,0) + \mu_2 p(0,1)$ -----------(1)

$(\lambda + \mu_1) p(1,0) = \mu_2 p(1,1) + \lambda p(0,0)$ ------(2)

$(\lambda + \mu_2) p(0,1) = \mu_1 p(1,1)$ -------------------(3)

$(\lambda + \mu_1 + \mu_2) p(1,1) = \mu_2 p(0,1) + \lambda p(1,0) + (\mu_1 + \mu_2) p(2,1)$ ------(4)

$(\lambda + \mu_1 + \mu_2)\, p(n,1) = \lambda\, p(n-1,1) + (\mu_1 + \mu_2)\, p(n+1,1)$, $n>1$ -----(5)

The traffic intensity of the system is :-

$\rho = \lambda / (\mu_1 + \mu_2)$

$0 = -(\lambda + \mu_k)\, p_k + \lambda\, p_{k-1} + p_{k+1}\, \mu_{k+1}$ , $k \geq 1$

Equation (5) is similar to the balanced equation of a birth–death process

Therefore,

$p(n,1) = \{ \lambda / (\mu_1 + \mu_2) \}\, p(n-1,1)$,    $n>1$ ------(6)

By repeated use of equation (6) we get :-

$p(n,1) = \rho \cdot p(n-1,1)$
$= \rho \cdot \rho \cdot p(n-2,1)$
::::::::::::::::::::
::::::::::::::::::::
$= \rho^{n-1} \cdot p(1,1)$ , $n>1$ -------------(7)

from equation (1) and (3) we can obtain by elimination :-

$$p(0,1) = \frac{\rho}{1+2\rho} \cdot \frac{\lambda}{\mu_2}\, p(0,0)$$

$$p(1,0) = \frac{1+\rho}{1+2\rho} \cdot \frac{\lambda}{\mu_1}\, p(0,0)$$

$$p(1,1) = \frac{\rho}{1+2\rho} \cdot \frac{\lambda(\lambda + \mu_2)}{\mu_1 \mu_2}\, p(0,0)$$

Some Simulated result and discussion

The figure -1 shows the propagation delay per jobs. propagation time is the time taken by the data to be transmitted
from source to destination. It is actually the delay taking place due to the physical medium. The graph does not show propagation time directly but measure the round trip time (RTT) as an estimate.

The nature of the graph as seen from the figure is somewhat spiky. We can see that the RTT is almost same for different jobs (by using the job of same size). However, it is not fixed and it can be seen from the figure that it can vary depending upon certain conditions. In the graph the average RTT is around one second. The best case is a

near –zero RTT and the worst case is slightly greater than two seconds.

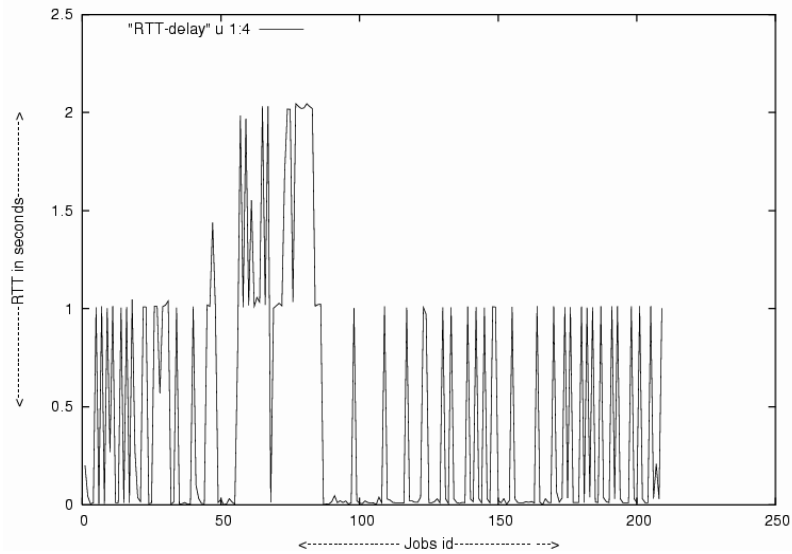

**Figure -1**

The figure -2 shows the amount of queuing delay for each individual job.
Initially the queue is empty and hence the incoming jobs do not have to wait much in the queue .But as time increases, more jobs come into the intermediate machine. The job arrival rate is much more than the rate at which the server the servers can process them. Hence the population in the queue increases. Since the jobs are sent to the servers following FCFS scheme, more number of jobs in the queue implies higher queuing delay for the jobs coming next. So if the clients throw jobs continuously (i.e in a particular session) the queuing delay is higher for the jobs that come later. When more than one job sending sessions are used, at the end of one session no new jobs arrives in the system for some time but the servers continue to pull jobs from the queue. Hence the number of jobs waiting in the queue decreases. Now at the beginning of the next session new jobs find lesser number of jobs waiting in the queue and they have to wait for lesser amount of time in the queue.

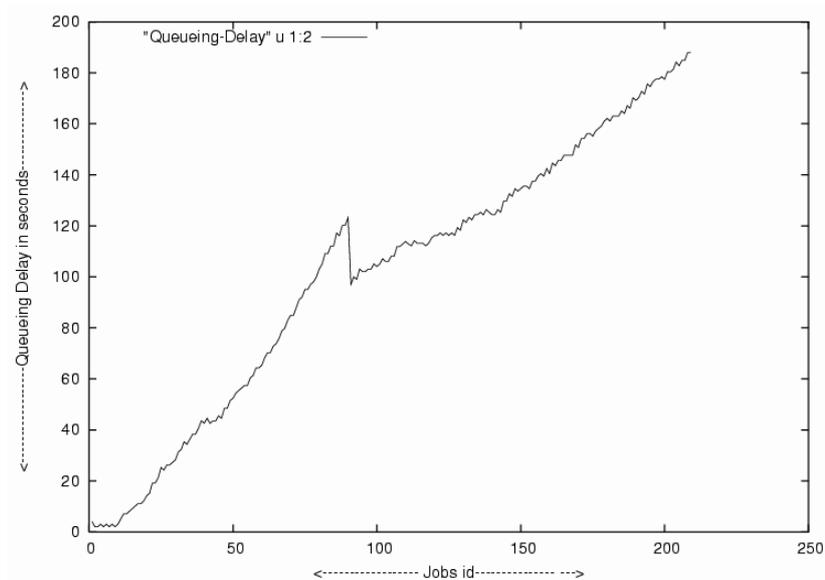

**Figure -2**

The figure-3 shows how much load or unprocessed jobs are waiting in the job queue at the intermediate machine in certain
Instance of time. The time axis here is started whenever the execution of dispatcher program begins.

The above graph shows the relationship between load vs time i.e the number of jobs waiting in queue at any instance. The rate of arrival of jobs from the clients is much more than the rate at which they can be dispatched to the two servers, hence as time increased load i.e the number of jobs queued up also increases. For example if 20 jobs are present at $t_1$ instant and 10 jobs arrive within $(t_1+1)^{th}$ instant then it may happen that only 5 jobs are dispatched in this interval i.e the load at $t_1$ was 20 and that at $t_1+1$ was 20+10-5=25. In the case when no jobs arrive, the load gradually decrease because the jobs are dispatched one by one without increasing the number of jobs in the queue. This explains the falling edge of the curve.

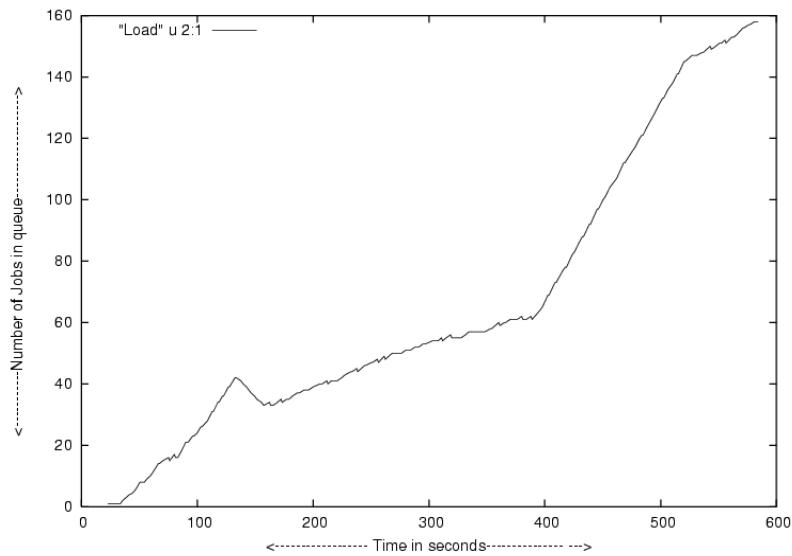

**Figure -3**

The figure -4 shows the queuing delay, it is time for the which a jobs waits in the queue before being served by the server and and by load it means the total number of jobs waiting in the queue at a particular instance.

The figure -4 shows, initially when the load was low the queuing delay was also minimum. Then as the load increase the queuing delay also increased. This is because initially the jobs didn't have to wait in the queue; they get serviced as they arrive in the queue.

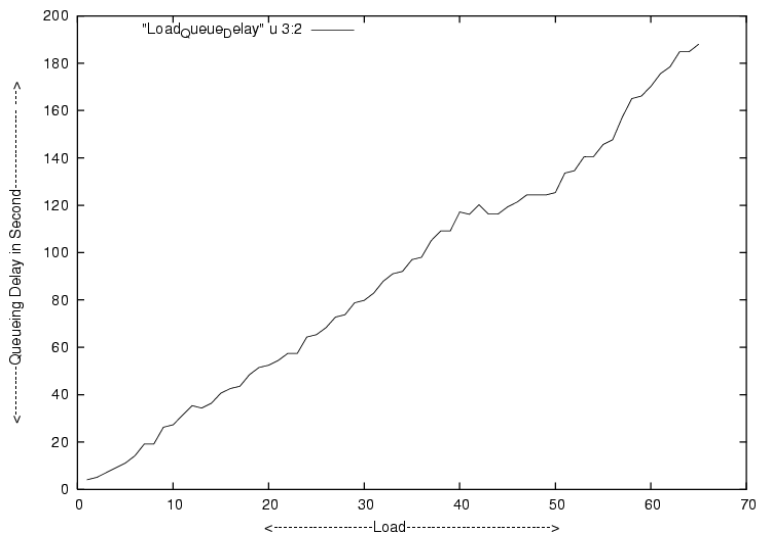

**Figure -4**

The figure -5 shows the traffic intensity $\rho = (\lambda / \mu)$ along Y –axis. It is defined as the ratio of job arrival rate ($\lambda$) and Job service rate ($\mu$). The X- axis the time elapsed since beginning of the execution of the program at the intermediate machine. The very beginning the servers are sitting idle, as there are no job assigned to service. Hence the service rate $\mu$ is zero thus making $\rho -> \infty$. for the simulation considering this value to high enough for this case so the curve at the beginning laying at very high value when jobs arrive at the intermediate machine they forwarded to the servers to get processed .so $\mu$ gets some value as well as $\lambda$ thus curve down to certain value depending on the intensity.

Whenever servers sitting idle, the traffic intensity $\rho$ has very high value to give a curve at the high end (as denominator $\mu$ becomes zero as explained earlier ) .And when servers are servicing previous jobs but queue has no jobs waiting (making $\lambda$ ->0) The curve lies at the X-axis ( i.e $\rho -> 0$ )

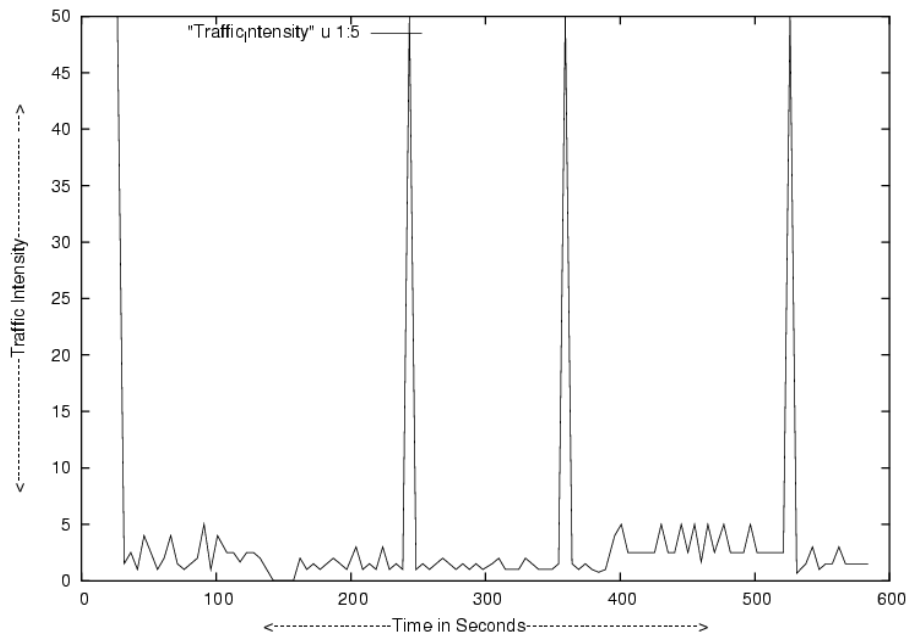

**Figure -5**

For the future, we want to integrate QoS and load balancing considerations in this problem to provide a comprehensive tree management architecture. I believe that such an architecture will be necessary to meet the requirements of future networks for high performance and reliability.

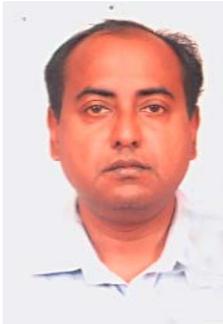
**Soumen Kanrar** received the M.Tech . degree in Computer Science and Data Processing from Indian Institute of Technology Kharagpur –India in 1999,and MS degree in Applied Mathematics From Jadavpur University – India in 1996 . During 2000 - 2008, he stayed in Durgapur Institute of Advance Technology as Academic Fellow .Currently working as a Research Associate in College of Computer & Information Sciences King Saud University, Riyadh.

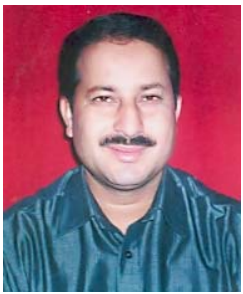
**Mohammad Siraj** received his M.E. degree in Computer Technology and Applications from Delhi College of Engineering, Delhi India in 1997,and BE degree in Electronics and Communication Engg.from Jamia Millia Islamia, New Delhi India in 1995. He has worked as Scientist in Defence Research and Development Organisation ,INDIA from 1995 -2000. Currently he is working as a Lecturer in College of Computer & Information Sciences King Saud University, Riyadh.